\documentclass[conference]{IEEEtran}

\usepackage{amsmath,amssymb}
\usepackage{multirow}
\usepackage[table,xcdraw]{xcolor}
\usepackage[utf8]{inputenc}
\usepackage{cleveref}
\crefname{section}{§}{§§}
\Crefname{section}{§}{§§}
\usepackage{float}
\usepackage{graphicx}
\usepackage{hyperref}
\usepackage{color}
\hypersetup{
	colorlinks   = true,
	citecolor    = blue,
	urlcolor     = blue
}
\usepackage{enumitem}
\makeatletter
\DeclareUrlCommand\ULurl@@{%
	\def{\scriptsize ($ \pm 0.00 $)}Font{\ttfamily\color{blue}}%
	\def{\scriptsize ($ \pm 0.00 $)}Left{\uline\bgroup}%
	\def{\scriptsize ($ \pm 0.00 $)}Right{\egroup}}
\def\ULurl@#1{\hyper@linkurl{\ULurl@@{#1}}{#1}}
\DeclareRobustCommand*\ULurl{\hyper@normalise\ULurl@}
\makeatother

\usepackage{array}
\newcolumntype{L}[1]{>{\raggedright\let\newline\\\arraybackslash\hspace{0pt}}m{#1}}
\newcolumntype{C}[1]{>{\centering\let\newline\\\arraybackslash\hspace{0pt}}m{#1}}
\newcolumntype{R}[1]{>{\raggedleft\let\newline\\\arraybackslash\hspace{0pt}}m{#1}}

\usepackage{caption}
\usepackage{subcaption}
\usepackage{amsfonts}
\usepackage{fixltx2e}
\usepackage{bm}
\usepackage{amsmath}
\usepackage{graphicx}
\usepackage{subcaption}
\usepackage{amssymb}
\usepackage{tikz}
\usepackage{array}
\usepackage{xcolor}
\usepackage[mathscr]{euscript}
\usepackage{mathrsfs}

\usetikzlibrary{shapes.geometric, arrows}
\tikzstyle{ADG} = [ellipse, minimum width=1cm, minimum height=.75cm,text centered, draw=black, fill=pink!30]
\tikzstyle{PDG} = [ellipse, minimum width=1cm, minimum height=.75cm,text centered, draw=black, fill=yellow!20]
\tikzstyle{SSCFP} = [ellipse, minimum width=1cm, minimum height=.75cm,text centered, draw=black, fill=green!10]
\tikzstyle{Ins} = [ellipse, minimum width=1cm, minimum height=.75cm,text centered, draw=black, fill=brown!10]
\tikzstyle{Signs} = [ellipse, minimum width=1cm, minimum height=.75cm,text centered, draw=black, fill=orange!10]

\usepackage{listings}
\lstdefinestyle{customc}{
	belowcaptionskip=1\baselineskip,
	breaklines=true,
	frame=L,
	xleftmargin=\parindent,
	language=Java,
	showstringspaces=false,
	basicstyle=\footnotesize\ttfamily,
	keywordstyle=\bfseries\color{green!40!black},
	commentstyle=\itshape\color{purple!40!black},
	identifierstyle=\color{blue},
	stringstyle=\color{orange},
}

\usepackage{pifont}
\usepackage {algorithm,algorithmicx}
\usepackage{algpseudocode}

\usepackage{verbatim}

\usepackage{enumerate}
\usepackage{enumitem}
\usepackage[space]{cite}
\usepackage{footmisc}
\usepackage{footnote}
\usepackage[nodisplayskipstretch]{setspace}

\makeatletter
\def\therule{\makebox[\algorithmicindent][l]{\hspace*{.5em}\vrule height .75\baselineskip depth .25\baselineskip}}%

\newtoks\therules
\therules={}
\def\appendto#1#2{\expandafter#1\expandafter{\the#1#2}}
\def\gobblefirst#1{
	#1\expandafter\expandafter\expandafter{\expandafter\@gobble\the#1}}%
\def\LState{\State\unskip\the\therules}
\def\pushindent{\appendto\therules\therule}%
\def\popindent{\gobblefirst\therules}%
\def\printindent{\unskip\the\therules}%
\def\printandpush{\printindent\pushindent}%
\def\popandprint{\popindent\printindent}%

\algdef{SE}[WHILE]{While}{EndWhile}[1]
{\printandpush\algorithmicwhile\ #1\ \algorithmicdo}
{\popandprint\algorithmicend\ \algorithmicwhile}%
\algdef{SE}[FOR]{For}{EndFor}[1]
{\printandpush\algorithmicfor\ #1\ \algorithmicdo}
{\popandprint\algorithmicend\ \algorithmicfor}%
\algdef{S}[FOR]{ForAll}[1]
{\printindent\algorithmicforall\ #1\ \algorithmicdo}%
\algdef{SE}[LOOP]{Loop}{EndLoop}
{\printandpush\algorithmicloop}
{\popandprint\algorithmicend\ \algorithmicloop}%
\algdef{SE}[REPEAT]{Repeat}{Until}
{\printandpush\algorithmicrepeat}[1]
{\popandprint\algorithmicuntil\ #1}%
\algdef{SE}[IF]{If}{EndIf}[1]
{\printandpush\algorithmicif\ #1\ \algorithmicthen}
{\popandprint\algorithmicend\ \algorithmicif}%
\algdef{C}[IF]{IF}{ElsIf}[1]
{\popandprint\pushindent\algorithmicelse\ \algorithmicif\ #1\ \algorithmicthen}%
\algdef{Ce}[ELSE]{IF}{Else}{EndIf}
{\popandprint\pushindent\algorithmicelse}%
\algdef{SE}[PROCEDURE]{Procedure}{EndProcedure}[2]
{\printandpush\algorithmicprocedure\ \textproc{#1}\ifthenelse{\equal{#2}{}}{}{(#2)}}%
{\popandprint\algorithmicend\ \algorithmicprocedure}%
\algdef{SE}[FUNCTION]{Function}{EndFunction}[2]
{\printandpush\algorithmicfunction\ \textproc{#1}\ifthenelse{\equal{#2}{}}{}{(#2)}}%
{\popandprint\algorithmicend\ \algorithmicfunction}%
\makeatother

\begin{document}

\title{Contextual Weisfeiler-Lehman Graph Kernel For Malware Detection}
\author{Annamalai Narayanan, Guozhu Meng, Liu Yang, Jinliang Liu and Lihui~Chen
	\\ Nanyang Technological University, Singapore.
	\\annamala002@e.ntu.edu.sg, \{gzmeng, yangliu\}@ntu.edu.sg,  liuj0081@e.ntu.edu.sg, elhchen@ntu.edu.sg}
\maketitle

\begin{abstract}
In this paper, we propose a novel graph kernel specifically to address a challenging problem in the field of cyber-security, namely, malware detection.
Previous research has revealed the following: (1) Graph representations of programs are ideally suited for malware detection as they are robust against several attacks, (2) Besides capturing topological neighbourhoods (i.e., structural information) from these graphs it is important to capture the context under which the neighbourhoods are reachable to accurately detect malicious neighbourhoods. 

We observe that state-of-the-art graph kernels, such as Weisfeiler-Lehman kernel (WLK) capture the structural information well but fail to capture contextual information. To address this, we develop the Contextual Weisfeiler-Lehman kernel (CWLK) which is capable of capturing both these types of information. 
We show that for the malware detection problem, CWLK is more expressive and hence more accurate than WLK while maintaining comparable efficiency. Through our large-scale experiments with more than 50,000 real-world Android apps, we demonstrate that CWLK outperforms two state-of-the-art graph kernels (including WLK)  and three malware detection techniques by more than 5.27\% and 4.87\% F-measure, respectively, while maintaining high efficiency. This high accuracy and efficiency make CWLK suitable for large-scale real-world malware detection.

\end{abstract}

keywords | graph kernels, malware detection, program analysis

\IEEEpeerreviewmaketitle

\vspace{-2mm}
\section{Introduction}
\label{sec:intro}
Malware detection has evolved as one of the challenging problems in the field of cyber-security as the attackers continuously enhance the sophistication of malware to evade novel detection techniques. Malware for various platforms such as desktop and mobile devices is growing at an alarming rate. For instance, Kaspersky reports \cite{Kas} detecting 4 million malware infections in 2015 which is a 216\% increase over 2014. This volume and growth rate clearly highlights an imperative need for automated malware detection solutions.\\
\nolinebreak
To perform automated malware detection, security analysts resort to program analysis and machine learning (ML) techniques.
Typically, this process involves extracting semantic features from suitable representations of programs (e.g., assembly code, call graphs) and detecting 
malicious code or behavior patterns using ML classifiers \cite{AppContext,Adagio,Drebin,CSBD,MLMalDetect,DroidSift}.\\
\nolinebreak
A major reason for such tremendous growth rate in malware is the production of \textit{malware variants}. Typically, the attackers produce large number of \textit{variants} of the same malware by resorting to techniques such as variable renaming and junk code insertion \cite{Drebin,Adagio,DroidSift,s&p}. These variants perform same malicious functionality, with apparently different syntax, thus evading syntax-based detectors. However, higher level semantic representations such as call graphs, control- and data-flow graphs, control-, data- and program-dependency graphs mostly stay similar even when the code is considerably altered \cite{Adagio,CSBD,DroidSift,s&p}. 
In this work, we use a common term, \textit{Program Representation Graph} (PRG) to refer to any of these aforementioned graphs.
As PRGs are resilient against variants, many works in the past have used them to perform malware detection. In essence, such works cast malware detection as a \textit{graph classification problem} and apply existing graph mining and classification techniques \cite{s&p}.
Some methods such as \cite{AppContext,Adagio,CSBD} note that ML classifiers are readily applicable on data represented as vectors and attempt to encode PRGs as feature vectors. Typically, these techniques face two challenges:
\nolinebreak
\begin{itemize}[leftmargin=*]
	\setlength\itemsep{0em}
\item \textbf{(C1) Expressiveness.} PRGs are complex and expressive data structures that characterize topological relationships among program entities. Representing them as vectors is a non-trivial task. In many cases vectorial representations of PRGs fail to capture all the vital information. For instance, AppContext \cite{AppContext}, a well-known Android malware detection approach represents apps as PRGs and ends up capturing features from individual nodes without their topological neighbourhood information. With such loss of expressiveness, attacks that span across multiple PRG nodes could not be effectively detected. 
\item \textbf{(C2) Efficiency.} The scale of malware detection problem is such that we have millions of samples already and thousands streaming in every day. 
Many classic graph mining based approaches (e.g., \cite{s&p}) are NP hard and have severe scalability issues, making them impractical for real-world malware detection \cite{Adagio,WinMalGK}.
\end{itemize}
\begin{figure*}[ht!]
	\centering
	\includegraphics[height=2.5cm,width=18cm]{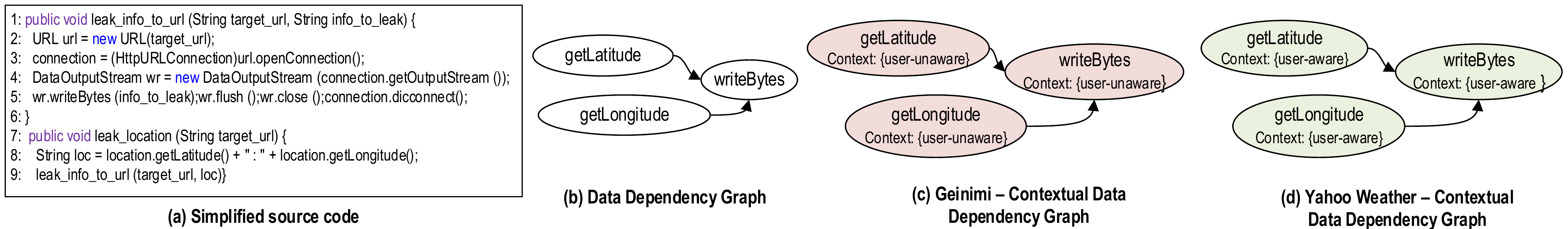}
	\caption{\small Location information being leaked in \textit{Geinimi} (malware) and \textit{Yahoo Weather} (benign) apps. (a) code snippet corresponding to leaking location information in both the apps. (b) DDG corresponding to the location leak. (c) \textit{Geinimi's} DDG illustrating that it leaks information without the user's knowledge. (d) \textit{Yahoo Weather's} DDG illustrating that it leaks information with the user's knowledge. \label {fig:me}}
\end{figure*}
\textbf{Graph Kernels.} One of the increasingly popular approaches in ML for graph-structured data is the use of graph kernels. 
Recently, efficient and expressive graph kernels such as \cite{WLK,NHGK,ContextualKernel,NSPDK} have been proposed and widely adopted in many application areas (e.g, bio- and chemo-informatics \cite{BioGK,CVGK,OptAlign,Molec}). 
Some of them support explicit feature vector representations of graphs (e.g, \cite{WLK,NHGK,ContextualKernel}). 
Thus both the aforementioned challenges C1 and C2 are effectively addressed by these graph kernels. Therefore, it just suffices to use a graph kernel together with a kernelized ML classifier (e.g., SVM) and we have a scalable, effective and ready-to-use malware detector. Recently, three approaches \cite{Adagio}, \cite{MLMalDetect} and \cite{WinMalGK}, \ have successfully demonstrated using these general purpose graph kernels for malware detection.\\
\nolinebreak
\textbf{Research Gap.} However, a major problem in using these general purpose graph kernels on PRGs is that, they are not designed to take domain-specific observations into account. 
For instance, recent research on malware analysis has revealed that besides capturing neighbourhood (i.e., structural) information from PRGs it is important to capture the context under which the neighbourhoods are reachable to accurately detect malicious neighbourhoods \cite{AppContext,DroidSift} (explained in detail in \S \ref{sec:bgm}). Many existing graph kernels such as \cite{WLK} and \cite{NHGK} can capture and compare structural information from PRGs effectively. However, they are not designed to capture the reachability context, as it is a strong domain-specific requirement and hence fail to do so. To address this, we develop a novel graph kernel which is capable of capturing both the aforementioned types of information. \\
\nolinebreak
For similar domain-specific reasons, researchers from other fields such as computer vision \cite{CVGK}, bio- and chemo-informatics \cite{BioGK,OptAlign,Molec} have developed a number of kernels that specifically suit their applications. 
Despite graphs being natural representations of programs and amenable for various activities, the program analysis research community has not devoted significant attention to development of domain-specific graph kernels. We take the first step towards this, by developing a kernel on PRGs which specifically suits our task of malware detection.\\
\nolinebreak
\textbf{Our approach.}  
To improve the accuracy of malware detection process, we propose a method to enrich the feature space of a graph kernel that inherently captures structural information with contextual information.
We apply this feature-enrichment idea on a state-of-the-art graph kernel, namely, Weisfeiler-Lehman kernel (WLK) \cite{WLK} to obtain the Contextual Weisfeiler-Lehman kernel (CWLK).
Specifically, CWLK associates to each sub-structure feature of WLK a piece of information about the context under which the sub-structure is reachable in the course of execution of the program. A sub-structure appearing in two different PRGs will match only if it is reachable under the same context in both PRGs. We show that for the malware detection problem, CWLK is more expressive and hence more accurate than WLK and other state-of-the-art kernels while maintaining comparable efficiency.  \\
\nolinebreak
\textbf{Experiments.} Through our large-scale experiments with more than 50,000 Android apps, we demonstrate that CWLK outperforms two state-of-the-art graph kernels (including WLK)  and three malware detection techniques by more than 5.27\% and 4.87\% F-measure, respectively, while maintaining high efficiency. This, in essence shows the significance of incorporating the contextual information along with structural information in the graph kernel while performing malware detection.\\
\textbf{Contributions.} The paper makes the following contributions:\\
(1)	We develop a graph kernel that captures both structural and contextual information from PRGs to perform accurate and scalable malware detection (\S\ref{sec:cwlk}). To the best of our knowledge, this is the first graph kernel specifically addressing a problem from the field of program analysis. \\
(2)	Through large-scale experiments and comparative analysis, we show that the proposed kernel outperforms two state-of-the-art graph kernels and three malware detection solutions in terms of accuracy, while maintaining high efficiency (\S\ref{sec:eval}).\\
(3)	We make an efficient implementation of the proposed kernel (along with the dataset information) publicly available\footnote{{https://sites.google.com/site/cwlkernel}}.\\
\nolinebreak
%
%
%
\vspace{-2mm}
\section{Background \& Motivation}
\label{sec:bgm}
In this section, we motivate the design of our kernel by describing why considering just the structural information from PRGs is insufficient to determine the maliciousness of a sample and how supplementing it with contextual information helps to increase detection accuracy. To this end we use a real-world Android malware\footnote{\scriptsize SHA256:05620032f3a2abd5ebea482b5e5d5b8ff5faa8115019736013d87f442032b6bc} from the \textit{Geinimi} family which steals users’ private information. We contrast its behavior with that of a well-known benign app, \textit{Yahoo Weather}.\\
\nolinebreak
\textbf{\textit{Geinimi}’s execution.} The app is launched through a background event such as receiving a SMS or call. Once launched, it reads the user’s personal information such as geographic location and contacts and leaks the same to a remote server. The (simplified) malicious code portion pertaining to the location information leak is shown in Fig. \ref{fig:me} (a). The method \textit{leak\_location} reads the geographic location through {\tt getLatitude} and {\tt getLongitude} Application Programming Interfaces (APIs). Subsequently, it calls \textit{leak\_info\_to\_url} method to leak the location details (through {\tt DataOutputStream.writeBytes}) to a specific server. The Data Dependency Graph (DDG) corresponding to the code snippet is shown in Fig. \ref{fig:me} (b). The nodes in DDG are labeled with the sensitive APIs that they invoke. \\
\nolinebreak
\textbf{\textit{Yahoo Weather}’s execution.} On the other hand, \textit{Yahoo Weather} could be launched only by user’s interaction with the device (e.g., by clicking the app’s icon on the dash board). The app then reads the user’s location and sends the same to its weather server to retrieve location-specific weather predictions. Hence, DDG portions of \textit{Yahoo Weather} is same as that of \textit{Geinimi}. \\
\nolinebreak
\textbf{Contextual information.} 
From the explanations above, it is clear that both the apps leak the same information in the same fashion. However, what makes \textit{Geinimi} malicious is the fact that its leak happens without the user’s consent. In other words, unlike \textit{Yahoo Weather}, \textit{Geinimi} leaks private information through an event which is not triggered by user’s interaction. We refer to this as a leak happening in \textit{user-unaware} context. On the same lines, we refer to \textit{Yahoo Weather’s} leak as happening in \textit{user-aware} context. \\
\nolinebreak
As explained in \cite{DroidSift} and \cite{AppContext}, in the case of Android apps, one could determine whether a PRG node is reachable under \textit{user-aware} or \textit{user-unaware} context by examining its entry point nodes. Following this procedure we add the context as an attribute to every DDG node. This context annotated DDG of \textit{Geinimi} and \textit{Yahoo Weather} are shown in Fig. \ref{fig:me} (c) and (d), respectively. \\
\nolinebreak
\textbf{Requirements for effective detection.} From the aforementioned example the two key requirements that makes a malware detection process effective can be identified: \\
\textbf{(R1) \textit{Capturing structural information.}} Since malicious behaviors often span across multiple nodes in PRGs, just considering individual nodes (and their attributes) in isolation is not enough. Capturing the structural (i.e., neighborhood) information from PRGs is of paramount importance. \\
\textbf{(R2) \textit{Capturing contextual information.}} Considering just the structural information without the context is not enough to determine whether a sensitive behavior is triggered with or without user’s knowledge. For instance, if structural information alone is considered, the features of both \textit{Geinimi} and \textit{Yahoo Weather} apps become identical, thus making the latter a false positive.  Hence, it is important for the detection process to capture the contextual information as well to make the detection process more accurate. \\
\nolinebreak
Many existing graph kernels could address the first requirement well. However, the second requirement which is more domain-specific makes the problem particularly challenging.  
To the best of our knowledge, none of the existing graph kernels support capturing this reachability context information along with structural information. 
Hence, this gives us a clear motivation to develop a new kernel that specifically addresses our two-fold requirement.  

\section{Definitions and Notations}
\label{sec:defnot}
\noindent
The formal definitions and notations that will be used throughout the paper are presented in this section. \\
\textbf{Definition 1 (Program Representation Graph).} $PRG = (N, E, \lambda, \xi)$ is a directed graph where $N$ is a set of nodes and each node $n \in N$ denotes program entity such as a function or instruction. $E \subseteq (N\times N) $ is a set of edges and each edge $e(n_1, n_2) \in E$ denotes either control- or data-flow or dependency from $n_1$ to $n_2$. $\lambda$ is the set of labels that characterize the (security-sensitive) operations of a node and $\ell: N \rightarrow \lambda$, is a labeling function which assigns a label to each node. $\xi$ is a set of events that denote the context of a node and $\mathcal{C}: N \rightarrow \xi$, is a function which assigns the context to each node.\\
\textbf{Definition 2 (Context).} The context of a node $ n \in N $ in the PRG of a program $P$ is a set of attributes $ \xi(n) $ that govern the reachability of $n$ in the course of execution of $P$.\\
\textbf{Examples of contexts.} 
In the case of Windows executables, the \textit{guard conditions} that govern the execution of a node could be considered as its context \cite{s&p}. Unlike Windows (and other desktop OS) binaries, Android and iOS mobile apps typically have multiple entry points \cite{DroidSift,AppContext}. Hence, in the case of such mobile apps, besides guard conditions, the categories of entry points through which a node is reachable could also be considered as its context. 
Similar platform-specific constraints and observations could be considered while defining the contexts for executables of other platforms.

\section {Contextual Weisfeiler-Lehman Graph Kernel}
\label{sec:cwlk}
In this section, we begin by explaining how the regular WLK can be applied to perform malware detection using PRGs and how it falls short. Subsequently, we introduce our CWLK and discuss how it addresses the shortcomings of WLK. Finally, we prove CWLK’s semi-definitiveness and analyze its time complexity.
\subsection{Regular Weisfeiler-Lehman Graph Kernel}
\label{subsec:wlk}
WLK computes the similarities between graphs based on the 1-dimensional WL test of graph isomorphism \cite{WLK}.\\ 
\textbf{WL test of isomorphism.} Suppose we are to determine whether a given a pair of graphs $G$ and $G'$ are isomorphic. The WL test of isomorphism works by augmenting the node labels by the sorted set of labels of neighboring nodes. 
This process is referred to as \textit{label-enrichment} and new labels are referred as \textit{neighborhood labels}. 
Thus, in each iteration \textit{i} of the WL algorithm, for each node $n \in N$, we get a new neighborhood label, $\lambda_{i}(n)$ that encompass the $i^{th}$ degree neighborhood around $n$. 
$\lambda_{i}(n)$ could be optionally compressed using a hash function $f: \Sigma^* \rightarrow \Sigma$ such that $f(\lambda_{i}(n)) = f(\lambda_{i}(n'))$, iff $\lambda_{i}(n) = \lambda_{i}(n')$. 
To test graph isomorphism, the re-labeling process is repeated until the neighborhood label sets of $G$ and $G'$ differ, or the number of iterations reaches a specific threshold.
Therefore, one iteration of WL relabeling is equivalent to a function $r((N,E,\lambda_i)) \! = \! (N,E,\lambda_{i+1})$  that transforms all graphs in the same manner. 
\\
\nolinebreak
\textbf{Definition 3 (WL sequence).} Define the WL graph at height $i$ of the graph $G = (N,E,\lambda)$ as the graph $\mathscr{G}_i = (V,E,\lambda_i)$. The sequence of graphs 
\begin{equation}
	{\mathscr{G}_0, \mathscr{G}_1, ..., \mathscr{G}_h} = {(V,E,\lambda_0),(V,E,\lambda_1), ...,(V,E,\lambda_h)}
\end{equation}
is called the WL sequence up to height $h$ of $G$, where $\mathscr{G}_0 = G$ (i.e., $\lambda_0 = \lambda$) is the original graph and $\mathscr{G}_1 = r(\mathscr{G}_0)$ is the graph resulting from the first relabeling, and so on. \\
\nolinebreak
\textbf{Definition 4 (WL kernel).} Given a valid kernel $k(.,.)$ and the WL sequence of graph of a pair of graphs $G$ and $G'$, the WL graph kernel with $h$ iterations is defined as 
\begin{equation}
\label{eq:wlk}
	k^{(h)}_{WL}(G, G') = k(\mathscr{G}_0, \mathscr{G}'_0) + ... + k(\mathscr{G}_h, \mathscr{G}'_h)
\end{equation}
where $h$ is the number of WL iterations and ${\mathscr{G}_0, \mathscr{G}_1, ..., \mathscr{G}_h}$ and ${\mathscr{G}'_0, \mathscr{G}'_1, ..., \mathscr{G}'_h}$ are the WL sequences of $G$ and $G'$, respectively. $ h $ is referred as \textit{height of the kernel}.\\
\nolinebreak
Intuitively, WLK counts the common neighborhood labels in two graphs. 
Hence we have $k^{(h)}_{WL}(G,G') = |(\lambda_i(n), \lambda_i(n'))|$, 
iff $ f(\lambda_i(n)) = f(\lambda_i(n'))$ for $i \in \{1, . . . , h\}, n \in N, n' \in N'$, where $f$ is injective and the sets $ \{f(\lambda_i(n))|n \in N\cup N'\} $ and $ \{f(\lambda_j(n))|n \in N\cup N'\} $ are disjoint for all $ i \neq j $.\\
\nolinebreak
\nolinebreak
\textbf{Example \& WLK's shortcoming.}
We now apply WLK on the real-world examples discussed in \S \ref{sec:bgm} to see if it distinguishes malicious and benign neighborhoods clearly, facilitating accurate detection. For the ease of illustration, the label compression step is avoided. Applying WLK on the DDG for both \textit{Geinimi} and \textit{Yahoo Weather} apps, shown in Fig. \ref{fig:me} (b), for the node {\tt getLatitude}, for heights $h = 0,1$, we get the neighborhood labels {\tt getLatitude} and {\tt getLatitude,writeBytes}, respectively. Clearly, WLK captures the neighborhood around the node {\tt getLatitude}, incrementally in every iteration of $h$. 
In fact, neighborhood label for $h=1$ captures that another sensitive node, {\tt writeBytes} lies in the neighborhood of {\tt getLatitude}, which highlights a possible privacy leak. However, WLK does not capture whether the neighborhood involved in this leak is reached in \textit{user-aware} or \textit{unaware} context. This is precisely what we address through our CWLK.


\subsection{Contextual Weisfeiler-Lehman graph Kernel}
\label{ss:cwlk}
\begin{algorithm}[t]
	\small
	\caption{CWLK - Contextual re-labeling}
	\textbf{Input}:\\
	$G = G_0 = (N, E, \lambda_0, \xi)$ | \textit {PRG} with set of nodes ($N$), set of edges ($E$) and set of node labels ($\lambda_0$) and context for each node ($\xi$)\\
	$h$ | number of iterations\\
	\textbf{Output}:\\
	$\{\mathcal{G}_0, \mathcal{G}_1,...,\mathcal{G}_h\}$ - contextual WL sequence of height $h$ 
	\begin{algorithmic}[1]
		
		\Procedure{Contextual Re-label}{$G,h$}
		\For {$i = $ 0 to $h$}
		\For {\textbf{all} $n \in N$}
		\LState $\sigma_i(n) \leftarrow \emptyset$
		\If {$i = 0$}
		\For {$c \in \xi(n)$}
		\LState $\sigma_i(n) \leftarrow \sigma_i(n) \cup c \oplus \lambda_0(n) $ 
		\EndFor
		\Else
		\LState $\mathcal{N}(n) \leftarrow \{m\ |\ (n,m) \in E\}$
		\LState $M_i(n) \leftarrow \{\lambda_{i-1}(m)\ |\ m \in \mathcal{N}(n) \}$ 
		\LState $\lambda_i(n) \leftarrow  \lambda_{i-1}(n) \oplus sort(M_i(n))$  
		\For {$c \in \xi(n)$}
		\LState $\sigma_i(n) \leftarrow \sigma_i(n) \cup c \oplus \lambda_i(n) $ 
		\EndFor
		
		\EndIf
		\LState $\sigma_i(n) \leftarrow join(\sigma_i(n))$ 
		\LState $\gamma_i(n) \leftarrow f(\sigma_i(n))$ 
		\EndFor
		\LState $\mathcal{G}_i \leftarrow (N,E,\gamma_i)$ 
		\EndFor
		\LState \textbf{return} \{$\mathcal{G}_0,\mathcal{G}_1,...,\mathcal{G}_h$\}
		\EndProcedure
		
	\end{algorithmic}
	\label{algo:cr}
\end{algorithm}
The goal of CWLK is to capture not only neighborhoods around the node, but also to include the contexts in which each of the neighborhoods is reachable in the PRG. To this end, we modify the re-labeling step of WLK so as to accommodate the context of every neighborhood. We refer to this process as \textit{contextual-relabeling} and the sequence of graphs thus obtained as \textit{contextual WL sequence}. \\
\nolinebreak
\textbf{Contextual re-labeling.}
Specifically, CWLK performs one additional step in the re-labeling process which is to attach the contexts of every node to its neighborhood label in every iteration. This in effect, indicates the contexts under which a particular neighborhood is reachable. The label thus obtained is referred to as \textit{contextual neighborhood label}. 
The contextual relabeling process is presented in detail in Algorithm \ref{algo:cr}.  \\
\nolinebreak
\nolinebreak
The inputs to the algorithm are PRG, $G$ and the degree of neighbourhoods to be considered for re-labeling, $h$. The output is the sequence of contextual WL graphs, $\{\mathcal{G}_0, \mathcal{G}_1,...,\mathcal{G}_h\}\!=\!\{(N,E,\gamma_0), (N,E,\gamma_1),...,(N,E,\gamma_h)\}$, where $\gamma_1,...,\gamma_h$ are constructed using the contextual relabeling procedure.\\
\nolinebreak
For the initial iteration $ i=0 $, no neighborhood information needs to be considered. Hence the contextual neighborhood label $ \gamma_0 (n) $ for all nodes $ n \in N $ is obtained by justing prefixing the contexts to the original node labels and compressing the same (lines 6-8,17-18). For $ i\!>\!0 $, the following procedure is used for contextual re-labeling.
Firstly, for a node $n \in N$, all of its neighboring nodes are obtained and stored in $\mathcal{N}(n)$ (line 10). For each node $m \in \mathcal{N}(n)$ the neighborhood label up to degree $i-1$ is obtained and stored in multiset $M_i(n)$ (line 11). $\lambda_{i-1}(n)$, neighborhood label of $n$ till degree $i\!-\!1$ is concatenated to the sorted value of $M_i(n)$ to obtain the current neighborhood label,  $\lambda_i(n)$ (line 12). Finally the current neighborhood label is prefixed with the contexts of node $n$ to obtain the string $ \sigma_i(n) $ which is then compressed using the function $ f $ to obtain the contextual neighborhood label,  $\gamma_i(n)$ (lines 13-15,17-18). \\
\nolinebreak
\textbf{Definition 5 (CWL kernel).} Given a valid kernel $k(.,.)$ and the CWL sequence of graph of a pair of graphs $G$ and $G'$, the contextual WL graph kernel with $h$ iterations is defined as 
\begin{equation}
\label{eq:cwlk}
k^{(h)}_{WL}(G, G') = k(\mathcal{G}_0, \mathcal{G}'_0) + ... + k(\mathcal{G}_h, \mathcal{G}'_h)
\end{equation}
where $h$ is the number of CWL iterations and ${\mathcal{G}_0, \mathcal{G}_1, ..., \mathcal{G}_h}$ and ${\mathcal{G}'_0, \mathcal{G}'_1, ..., \mathcal{G}'_h}$ are the CWL sequences of $G$ and $G'$, respectively. \\
\nolinebreak
Intuitively, CWLK counts the common contextual neighborhood labels in two graphs. 
Hence we have $k^{(h)}_{CWL}(G,G') = |(\sigma_i(n), \sigma_i(n'))|$, 
iff $ f(\sigma_i(n))\!=\!f(\sigma_i(n'))$ for $i \in \{1, . . . , h\}, n \in N, n' \in N'$.\\
\nolinebreak
\textbf{Example.}
We now apply CWLK on the apps in our example to show how it overcomes WLK's shortcomings. The contextual neighborhood labels $ \sigma $ (without compression) of the node {\tt getLatitude} in \textit{Geinimi} app for heights $ h\!=\!0,1$ are, {\tt \textbf{user-unaware}}$\oplus${\tt getLatitude} and {\tt \textbf{user-unaware}}$\oplus${\tt getLatitude,witeBytes}, respectively. For the same node in \textit{Yahoo Weather} the contextual neighborhood labels are {\tt \textbf{user-aware}}$\oplus${\tt getLatitude} and {\tt \textbf{user-aware}}$\oplus${\tt getLatitude,witeBytes}. Hence, it is evident that the CWLK’s contextual relabeling provides a means to clearly distinguish malicious PRG neighborhoods from the benign ones. This is achieved by complementing the structural information with contextual information. Therefore, unlike WLK, CWLK based classification does not detect \textit{Yahoo Weather} as a false positive. This example clearly establishes the suitability of CWLK for the malware detection task. \\
\noindent
We now prove CWLK's positive definiteness and also analyze its time complexity.\\
\textbf{Theorem 1.} CWLK is positive definite.\\
\textbf{Proof.} Let us define a mapping $\phi$ that counts the occurrences of a particular contextual neighborhood label sequence $\sigma$ in $G$ (generated in $h$ iterations of Algorithm \ref{algo:cr}). Let $\phi^{(h)}_\sigma(G)$ denote the number of occurrences of $\sigma$ in $G$, and analogously $\phi^{(h)}_\sigma(G')$ for $G'$. Then,
\begin{equation}
\begin{aligned}
	k^{(h)}_{\sigma}(G, G') = \phi^{(h)}_\sigma(G), \phi^{(h)}_\sigma(G') = |\{(\sigma_i(n),\sigma_i(n'))|\\ \sigma_i(n) = \sigma_i(n'), i \in \{1,...,h\}, n \in N, n' \in N'\}|
\end{aligned}
\end{equation}
\vspace{-2mm}
Summing over all $ \sigma $ from the vocabulary $ \Sigma^* $, we get
{\small \begin{multline}
k^{(h)}_{CWL}(G, G') = \sum_{\sigma \in \Sigma^*}^{} k^{(h)}_{\sigma}(G, G') = \sum_{\sigma \in \Sigma^*} \phi^{(h)}_\sigma(G)\phi^{(h)}_\sigma(G')\\ = |\{(\sigma_i(n),\sigma_i(n'))|\sigma_i(n) = \sigma_i(n'), i \in \{1,...,h\}, n \in N, n' \in N'\}|\\ = 
|\{(\sigma_i(n),\sigma_i(n'))|f(\sigma_i(n)) = f(\sigma_i(n')),\\ i \in \{1,...,h\}, n \in N, n' \in N'\}|
\end{multline}}
where the last equality follows from the fact that $ f $ is injective.\\
\nolinebreak
As $ f(\sigma) \neq f(\sigma')$ if $ \sigma \neq \sigma'$, the string $ \sigma $ corresponds to exactly one contextual neighborhood label and $ k^{(h)}_{CWL} $ defines a kernel with corresponding feature map $ \phi^{(h)}_{CWL} $, such that
\begin{equation}
 \phi^{(h)}_{CWL} = (\phi^{(h)}_{\sigma} (G))_{\sigma \in \Sigma^*} 
\end{equation}
\nolinebreak
\textbf{Complexity.} The runtime complexity of CWLK with $h$ iterations on a graph with $n$ nodes and $e$ edges is $O(he)$ (assuming that $ e\! > \! n $) which is same as that of WLK.
More specifically, the neighborhood label computation with sorting operations (lines 10-12  of Algorithm \ref{algo:cr}) take $ O(e) $ time for one iteration and the same for $ h $ iterations take $ O(he) $.
The inclusion of context (lines 6-8,13-15), does not incur additional overhead as $ e\! >>\! |\xi| $. Hence the final time complexity remains as $O(he)$. For a detailed derivation and analysis of the time complexity of WLK, we refer the reader to \cite{WLK}.\\
\nolinebreak
\textbf{Efficient computation of CWLK on K graphs.} When computing CWLK on $K$ graphs to obtain $K \times K$ kernel matrix, a naïve approach would involve $K^2$ comparisons, resulting a time complexity of $O(K^2he)$. However, as mentioned in \cite{WLK}, a Bag-of-Features (BoF) model based optimization could be performed to arrive the kernel matrix in $O(Khe+K^2hn)$ time. This optimized computation involves the following steps: (1) A vocabulary $\Sigma$ of all the contextual neighbourhood labels of nodes across the $K$ graphs is obtained in $O(Khe)$ time. This facilitates representing each of the $K$ graphs as feature vectors of $|\Sigma|$ dimensions. (2) Subsequently, $K \times K$ kernel matrix can be computed by multiplying these vectors in $O(K^2hn)$ time. \\
\nolinebreak
In summary, CWLK has the same efficiency as that of WLK and supports explicit feature vector representations of PRGs.\\
\nolinebreak
\textbf{Relation to other spatial contextual kernels.}
Two recently proposed graph kernels \cite{ContextualKernel} and \cite{NSPDK}, consider incorporating the spatial context information to neighborhood subgraph features. They define \textit{context of a subgraph feature as another subgraph appearing in its vicinity}.
As mentioned earlier, in our malware detection problem we refer to \textit{attributes of a node which determine its reachability as its context}. 
This \textit{reachability context} is different from \textit{spatial context} discussed in \cite{ContextualKernel} and \cite{NSPDK}. Hence CWLK is consummately different from these two kernels.
\begin{table}[ht]
	\centering
	\small
	\caption{Composition of dataset}
	\setlength\tabcolsep{3.25pt}
	\label{tab:DS}
	\begin{tabular}{|c|c|c|}
		\hline
		\textbf{Portion of dataset} & \textbf{Dataset source} & \textbf{\# of samples} \\ \hline
		Malware portion             & \textsc{Drebin} \cite{Drebin}, Virus Share \cite{VS}    & 29877                  \\ \hline
		Benign portion              & Google Play \cite{GP}             & 25000                  \\ \hline
	\end{tabular}
\end{table}
\vspace{-3mm}
\section {Evaluation}
\label{sec:eval}
We conducted large scale experiments involving more than 50,000 Android apps from two real-world malware datasets to evaluate the accuracy and efficiency of CWLK. We compare CWLK's performance against that of two state-of-the-art kernels and three Android malware detection solutions.

\subsection {Datasets}
\label{subsec:ds}
\textsc{Drebin} \cite{Drebin} provides a collection on 5,560 Android malware apps collected from 2010 to 2012. More recently, Virus-share \cite{VS} released a collection of 24,317 malware apps collected from 2010 to 2014. We combined these two datasets and use them in our evaluation. For the benign portion of the dataset, we collected 25,000 benign top-selling apps from Google Play store \cite{GP} that were released around the same time. Thus, our dataset contains a total of	54,877 apps. The composition of our dataset is presented in Table \ref{tab:DS}. 
\\
\nolinebreak
\textbf{Graph Representations considered.} As mentioned in \S \ref{sec:intro} the proposed CWLK could be applied on any type of PRGs to perform malware detection. 
In our evaluations, we experimented with two types of PRGs namely, call graphs (CGs) and inter-procedural control-flow graphs (ICFGs). The nodes of a CG represent the methods present in an app; its directed edges represent the \textit{calling/called relations} between the methods.
The nodes of an ICFG represent the individual instructions present in those methods; its edges represent the \textit{control-flows} among those instructions.
These two types of PRGs are chosen since they capture program semantics at different levels of granularities. 
The statistics on the average number of nodes and edges of the PRGs in our dataset are presented in Table \ref{tab:DSStat}. 
Details on construction of these PRGs are presented later in \S\ref{subsec:ed-impl}.
\begin{table}[t]
	\centering
	\scriptsize
	\caption{Dataset Statistics}
	\setlength\tabcolsep{6pt}
	\label{tab:DSStat}
	\begin{tabular}{|l|l|l|l|}
		\hline
		\multicolumn{2}{|c|}{\textbf{CG (Avg $\pm$ Std)}} & \multicolumn{2}{c|}{\textbf{ICFG (Avg $\pm$ Std)}} \\ \hline
		\textbf{\# of Nodes}    & \textbf{\# of Edges}    & \textbf{\# of Nodes}     & \textbf{\# of Edges}    \\ \hline
		1556 {\scriptsize $ \pm $ 998} &   3327 {\scriptsize $ \pm $ 1803} & 15323 {\scriptsize $ \pm $ 9844} & 22745 {\scriptsize $ \pm $ 20922}  \\ \hline
	\end{tabular}
\end{table}
\\
\noindent
\textbf{Training and Test sets.} 
60\% of the samples were randomly chosen from the datasets and used for training the classifier and the remaining 40\% samples are used to test their performances. 
The classifiers' hyper-parameters are determined on the training set using 5-fold cross-validation, whereas the test set is only used for determining the final detection performance. We repeat this procedure 5 times and
average the results.\\
\vspace{-3mm}
\subsection{Implementation \& Comparative Analysis}
\label{subsec:ed-impl}
\noindent
\textbf{PRG construction \& Context Identification.} Both CG and ICFG of the apps in our datasets are constructed through static analysis using \textit{Soot} \cite{Soot}, a well-known Android static analysis workbench. The nodes are labeled with the security-sensitive APIs they access and are annotated with context information. 
Nodes that do not access any sensitive APIs are removed. 
We use the category of entry points of each of the nodes in these graphs as their contexts. The procedure proposed in DroidSIFT \cite{DroidSift} is used to identify and categorize each entry point as being in ‘user-aware’ or ‘user-unaware’ context. \\
\nolinebreak
\textbf{Comparison with Graph Kernels.} CWLK's accuracy and efficiency is compared against those of WLK \cite{WLK} and Neighborhood
Hash Graph Kernel (NHGK) \cite{NHGK}.
Since these kernels cannot capture context information, we use the comparative analysis against them to ascertain whether including context information significantly improves the accuracy without affecting the efficiency. We implemented all these kernels in about 2,170 lines of Python code. For all the kernels, BoF model based implementation similar to the one discussed in \S \ref{ss:cwlk} is used obtain explicit feature vector representation of samples. \\
\nolinebreak
\textbf{Comparison with Malware Detection Solutions.} Also, our approach is compared against three light-weight state-of-the-art ML based Android malware detection solutions, namely, \textsc{Drebin} \cite{Drebin}, Allix \textit{et al.} \cite{CSBD} and \textsc{Adagio} \cite{Adagio}. To this end, we re-implemented \textsc{Drebin} and Allix \textit{et al.}'s approaches through consultations with the authors. For \textsc{Adagio}, an open-source implementation provided by the authors is used. 
\\
\nolinebreak
Since the malware detection accuracy of these solutions predominantly depend on the features they use, we briefly introduce them here. \textsc{Drebin} \cite{Drebin} considers features such as sensitive Android APIs, permissions and components used by apps. Allix \textit{et al.} \cite{CSBD} constructs the CFGs of individual methods and represent them as signature strings which are subsequently used as features. \textsc{Adagio} \cite{Adagio} constructs CGs and uses byte-code instructions to assign labels to nodes. NHGK \cite{NHGK} is used to extract CG neighborhoods as features and a histogram-intersection (HI) kernel SVM is trained to detect malware. Due to severe scalability issues (explained in \S \ref{subsec:comp}) \textsc{Adagio} is ran only once on our dataset. All other techniques are ran 5 times and the average results are reported. 

\begin{figure}[t]
	\centering
	\includegraphics[height=3.2cm,width=7.3cm]{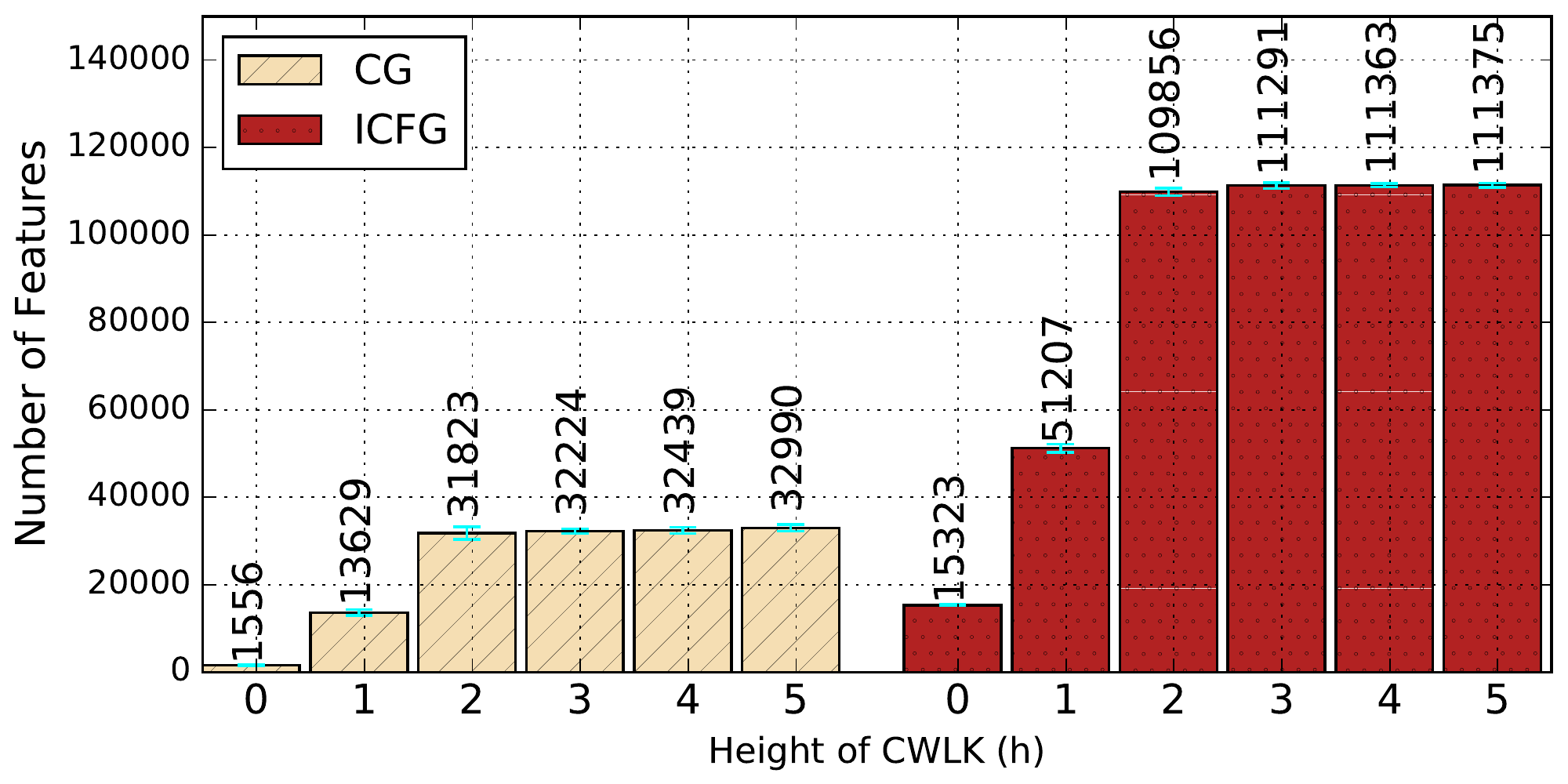}
	\caption{No. of features obtained for various heights of CWLK. \label {fig:feats}}
\end{figure}
\begin{table*}[t]
	\scriptsize
	\setlength\tabcolsep{3pt}
	\caption{Comparison of CWLK against state-of-the-art Graph Kernels (metrics expressed as \% values) - avg. over 5 runs}
	\label{tab:acc}
	\begin{tabular}{|c|l|l|l|l|l|l|l|l|l|l|}
		\hline
		\multirow{2}{*}{\textbf{\begin{tabular}[c]{@{}c@{}}PRG\\ Type\end{tabular}}} & \multirow{2}{*}{\textbf{Kernel}} & \multicolumn{3}{c|}{\textbf{\textit{h} = 0}}                                                                                      & \multicolumn{3}{c|}{\textbf{\textit{h} = 1}}                                                                                      & \multicolumn{3}{c|}{\textbf{\textit{h} = 2}}                                                                                      \\ \cline{3-11} 
		&                                  & \multicolumn{1}{c|}{\textbf{Precision}} & \multicolumn{1}{c|}{\textbf{Recall}} & \multicolumn{1}{c|}{\textbf{F-measure}} & \multicolumn{1}{c|}{\textbf{Precision}} & \multicolumn{1}{c|}{\textbf{Recall}} & \multicolumn{1}{c|}{\textbf{F-measure}} & \multicolumn{1}{c|}{\textbf{Precision}} & \multicolumn{1}{c|}{\textbf{Recall}} & \multicolumn{1}{c|}{\textbf{F-measure}} \\ \hline \hline
		\multirow{3}{*}{\textbf{CG}} & \textbf{WLK} \cite{WLK} & 71.21 ($ \pm 1.15 $) & \textbf{80.93} ($ \pm 1.78 $) & 75.75 ($ \pm 1.28 $) & 85.00 ($ \pm 1.59 $) & 85.93 ($ \pm 2.10 $) & 85.46 ($ \pm 1.39 $) & 87.28 ($ \pm 1.02 $) & 86.15 ($ \pm 0.78 $) & 86.71 ($ \pm 0.88 $) \\ 
		& \textbf{NHGK} \cite{NHGK} & 63.05 ($ \pm 3.11 $) & 60.14 ($ \pm 2.04 $) & 61.56 ($ \pm 2.56 $) & 67.18 ($ \pm 1.76 $) & 69.02 ($ \pm 2.01 $) & 68.08 ($ \pm 1.88 $) & 74.17 ($ \pm 1.11 $) & 70.22 ($ \pm 1.69 $) & 72.14 ($ \pm 1.88 $) \\ 
		& \textbf{CWLK} & \textbf{77.29} ($ \pm 1.75 $) & 79.85 ($ \pm 0.83 $) & \textbf{78.54} ($ \pm 0.97 $) & \textbf{88.50} ($ \pm 0.95 $) & \textbf{87.98} ($ \pm 1.41 $) & \textbf{88.23} ($ \pm 1.10 $) & \textbf{89.19} ($ \pm 0.83 $) & \textbf{90.10} ($ \pm 0.96 $) & \textbf{89.64} ($ \pm 0.94 $) \\ \hline
		\multirow{3}{*}{\textbf{ICFG}} & \textbf{WLK} \cite{WLK} & 85.49 ($ \pm 2.59 $) & \textbf{81.66} ($ \pm 3.71 $) & 83.53 ($ \pm 2.19 $) & 90.17 ($ \pm 1.92 $) & 87.48 ($ \pm 1.20 $) & 88.8 ($ \pm 1.37 $) & 93.09 ($ \pm 0.81 $) & 88.16 ($ \pm 0.77 $) & 90.55 ($ \pm 0.80 $) \\ 
		& \textbf{NHGK} \cite{NHGK} & 68.68 ($ \pm 4.10 $) & 70.73 ($ \pm 2.77 $) & 69.68 ($ \pm 3.20 $) & 75.78 ($ \pm 2.18 $) & 81.01 ($ \pm 2.64 $) & 78.30 ($ \pm 2.06 $) & 78.11 ($ \pm 1.07 $) & 82.28 ($ \pm 1.50 $) & 80.14 ($ \pm 1.29 $) \\ 
		& \textbf{CWLK} & \textbf{89.74} ($ \pm 2.33 $) & 81.25 ($ \pm 3.12 $) & \textbf{85.28} ($ \pm 2.59 $) & \textbf{96.77} ($ \pm 1.63 $) & \textbf{92.08} ($ \pm 1.20 $) & \textbf{94.36} ($ \pm 1.15 $) & \textbf{97.15} ($ \pm 0.28 $) & \textbf{94.53} ($ \pm 0.75 $) & \textbf{95.82} ($ \pm 0.66$)\\ \hline
	\end{tabular}
\end{table*}
\subsection{Experimental Design}
\noindent
\textbf{Research Questions.} Through our evaluations, we seek to address the following research questions: 
(1) Does including the context information in PRGs through CWLK significantly improve the malware detection accuracy? 
(2) Does capturing context information incur significant computation overhead to adversely affect the efficiency? 
(3) Does our context-based detection approach outperform existing malware detection solutions? 
Two separate experiments are designed to address these questions.\\
\nolinebreak
\textbf{Experiment E1.} In order to evaluate the first two questions, the following experiment is conducted: 
First, the CGs of all the training set apps are constructed. Then, the WLK, NHGK and CWLK kernels are applied on the CGs to obtain respective kernel matrices. Subsequently, a SVM classifier is trained with each of these kernels to detect malicious apps. Finally, the test set apps are subjected to the aforementioned kernel computation processes and are used to evaluate the models. The same procedure is repeated with ICFGs.\\
\textbf{Experiment E2.} In order to address question (3), we subject \textsc{Drebin} \cite{Drebin}, Allix \textit{et al.} \cite{CSBD} and \textsc{Adagio} \cite{Adagio} solutions to the same training and test sets. We compare them against the best performing model that uses CWLK (obtained from experiment E1).\\
\textbf{Setting the parameter \textit{h}:} We experimented with different kernel heights $h = $ 0 to 5 for CWLK (see eq. (3)). The average number of contextual neighborhood features (from 5 runs of E1) for different values of $ h $ on the two PRGs is reported in Fig. \ref{fig:feats}. It is evident that the number of features does not increase significantly after $h$ = 2 on both PRGs. This is because we have removed nodes that do not access sensitive APIs which affects the connectivity and restricts the neighborhood sizes. In other words, we seldom have neighborhoods around nodes that span beyond degree 2. Similar trend is observed in WLK and NHGK features. Hence we restrict the height $ h $ to be 0, 1 and 2 for all three kernels. Thus, for each  kernel applied on each PRG, we have three SVM classifiers (one for each $ h $). Therefore we have a total of 18 malware detection models under comparison (see Table \ref{tab:acc} for details). \\
\nolinebreak
\textbf{Evaluation metrics.} Standard evaluation metrics such as Precision, Recall and F-measure are used to determine the effectiveness of malware detection \cite{CSBD}. Efficiency is determined in terms of training and testing durations. \\
\nolinebreak
\textbf{Evaluation Setup.} All the experiments were conducted on Intel Xeon Hexa-core E5-2640 processor (2.50 GHz) with 32 GB RAM running Ubuntu 14.04.\\

\vspace{-3mm}
\subsection {Results and Discussions}
\label{subsec:rd}
\subsubsection {\textbf{Impact of context information on F-measure}}
\label{subsec:acc}
We compare the Precision, Recall and F-measure of the malware detection process through CWLK with those of WLK and NHGK. This is to ascertain whether incorporating the context information in PRGs boosts the effectiveness of detection. These results for the 18 models are presented in Table \ref{tab:acc}, from which the following inferences are drawn:
\begin{itemize}[leftmargin=*]
	\setlength\itemsep{0em}
	
	\item At the outset, two general observations are made: (1) All models perform better on ICFGs compared to CGs. This is because, ICFG is a more fine-grained representation of programs than CG, which enables it to capture program semantics at a finer level and thereby boosting the detecting accuracy. 
	Hence, we conclude that in our experiments ICFG is a more effective representation for malware detection. 
	(2) Considering larger neighborhoods helps capturing the structural information better which in turn reflects in better performances. This is evident as Precision, Recall and F-measure values get better with increasing values of $ h $ for all the three kernels. However, this observation may not hold for large values of $ h $, as nodes that are far apart will be considered for neighborhood re-labeling, leading to a noisy re-labeling process.
	
	\item It is clear that CWLK outperforms both WLK and NHGK in terms of F-measure consistently on both CG and ICFG representations.
	
	\item \textit{Since the only difference between WLK and CWLK is the latter’s capability to capture the context information, evidently this is the reason for CWLK’s superior performance}.
	
	\item Also, CWLK achieves better Precision than WLK in all the experiments, consistently. This indicates that CWLK suffers lesser false positives than WLK. This reduction in false positives is a direct result of capturing context information which helps to precisely distinguish malicious neighborhoods in the PRGs from the benign ones (as discussed in \S \ref{sec:cwlk}). 
	
	\item NHGK consistently fails to produce better results than WLK and CWLK. This is because, the node labeling and hashing technique adapted in NHGK causes collisions among informative and non-informative subgraph features.
    Such collisions could be avoided in WLK and CWLK through label compression (as explained in \S \ref{sec:cwlk}), making them more expressive and accurate. For further details on this limitation of NHGK, we refer the reader to the original work at \cite{NHGK} (particularly Section 5 where the authors discuss this limitation in detail). 
	
\end{itemize}
Since CWLK uses both contextual and structural information, it is important to analyze the contribution of each of these types of information to its performance. 
Capturing only structural information is equivalent to using WLK. Hence from rows 1 and 4 of Table \ref{tab:acc} (ignoring columns for $ h = 0 $), it is evident that structural information alone could provide a minimum of 85.46\% and an average of 87.88\% F-measure across both PRGs. 
Similarly, the contribution of contextual information alone is ascertained using CWLK and setting $h=0$ to be a minimum of 78.54\% and an average of 81.91\% F-measure. Finally, the effect of using both types of information is ascertained by using CWLK and setting $ h>0 $ to be a minimum of 88.23\% and an average of 92.02\% F-measure across both PRGs. This clearly conveys that structural information is primary for performing effective malware detection and contextual information complements it, thereby helping to improve the accuracy. CWLK attains superior performance by capturing both these types of information. 

%

\subsubsection {\textbf{Impact of context information on efficiency}}
\label{subsec:scale}
\begin{figure}[t]
	\includegraphics[height=4cm,width=9cm]{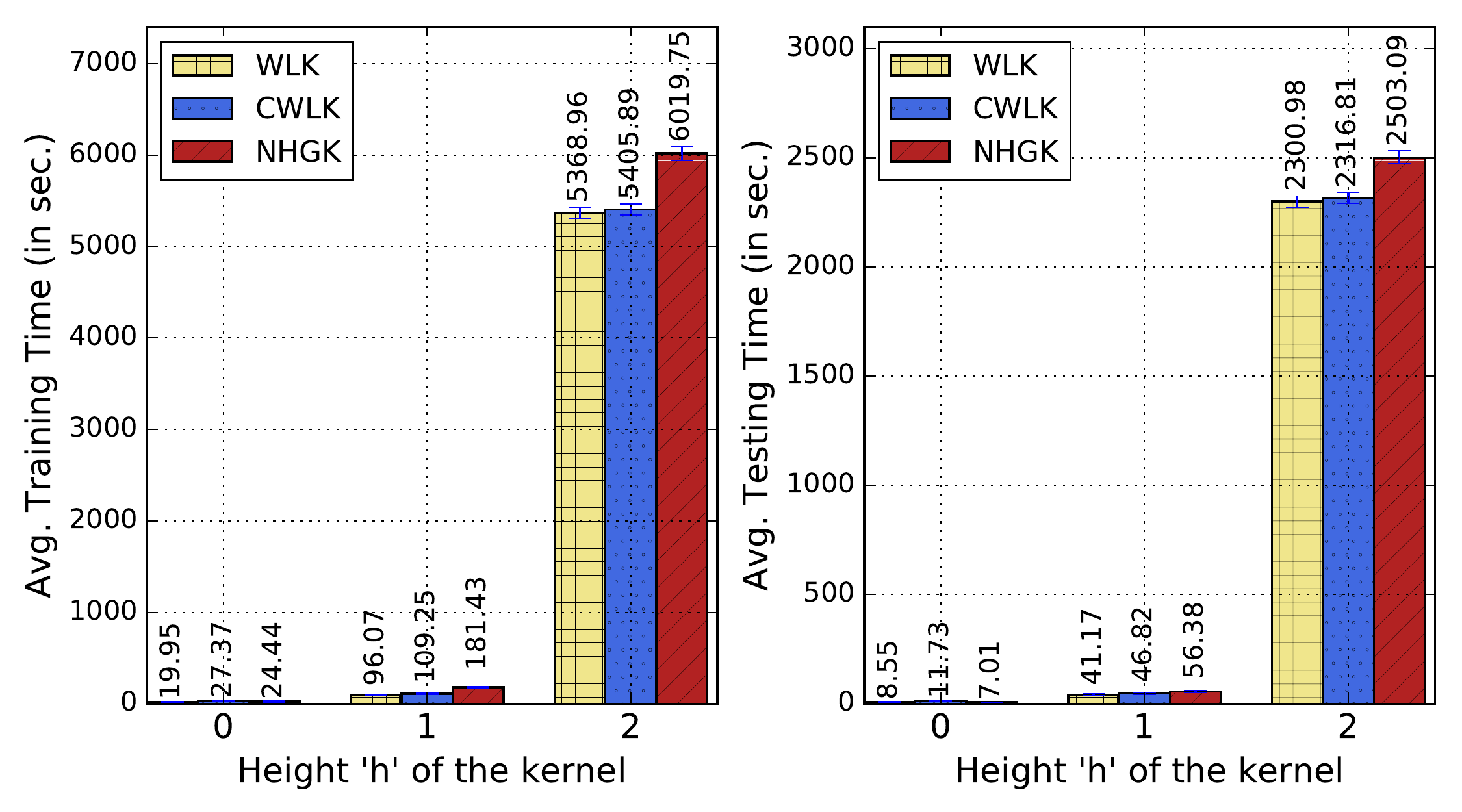}
	\caption{Training and Testing durations for different values of $h$ on ICFG representation (avg. over 5 runs) \label {fig:eff}}
\end{figure}
\noindent
We compare the training and testing durations of classifiers using the three kernels under study from experiment E1. This is to infer whether CWLK achieves higher accuracy at the cost of higher computation time. As discussed earlier in \S\ref{sec:cwlk}, CWLK’s time complexity is same as that of WLK. Hence, we expect negligible difference in efficiency. \\
\textit{Experimental Setting.} Due to space constraints, the training and testing times of the \textit{worst-case experimental setting} involving samples with maximum number of nodes and edges are only reported. This setting corresponds to using ICFG representations of apps.
The average training and testing durations in this experimental setting for various values of $ h $ over 5 runs of experiment E1 are presented in Fig. \ref{fig:eff}, from which the following inferences are drawn. 

\begin{itemize}[leftmargin=*]
	\setlength\itemsep{0em}

\item The training and testing durations of CWLK and WLK are almost same for all values of $h$. This is due to the fact that adding the context information which is the only additional operation in CWLK results in negligible difference on the total kernel computation, training and testing durations.

\item The training and testing durations increase with the height of the kernel $ h $ for all the kernels in a similar fashion. This is mainly due to the fact that both WLK and CWLK condense the neighborhood labels in the same way which is very similar to that of NHGK. Also, for all these kernels, there is a huge increase in training and testing durations when $ h $ increases from 1 to 2 which does not reflect proportionally in terms of F-measure. Hence, one may choose to perform malware detection in a much scalable and reasonably accurate manner by setting $ h=1 $ (i.e., considering only degree-1 neighbors).
 
\item It is noted that CWLK along with the other two kernels in Fig. \ref{fig:eff}, shows high efficiency as it operates in linear time on the density of PRGs. In particular, it is more efficient than the classic walk-, tree- and path-based graph kernels (discussed in \cite{GKVishy}) and is suitable for large-scale malware detection.
\end{itemize}
\noindent
\textbf{Summary.} From experiment E1, we conclude that CWLK through its virtue of capturing both structural and contextual information, significantly improves upon the accuracy of WLK for the malware detection task without hurting the efficiency.

\begin{table}[t]
	\setlength\tabcolsep{5pt}
	\scriptsize
	\centering
	\caption{Comparison of CWLK based malware detection against the state-of-the-art detectors (avg. over 5 runs)}
	\label{tab:maldetect}
	\begin{tabular}{|c|c|c|c|}
		\hline
		\textbf{Method}      & \textbf{Precision}            & \textbf{Recall}            & \textbf{F-measure}           \\ \hline \hline
		\textbf{\textsc{Drebin}} \cite{Drebin}      & 97.02 {\scriptsize ($ \pm $0.93)} & 85.60 {\scriptsize ($ \pm $1.03)}          & 90.95 {\scriptsize ($ \pm $0.71)}          \\ \hline
		\textbf{Allix \textit{et al.}} \cite{CSBD} & 88.24 {\scriptsize ($ \pm $0.74)}          & 86.29 {\scriptsize ($ \pm $1.66)} & 87.25 {\scriptsize ($ \pm $0.54)}          \\ \hline
		\textbf{\textsc{Adagio}} \cite{Adagio}      & 92.18                 & 87.32                 & 89.68                 \\ \hline
		\textbf{CWLK}        & \textbf{97.15 {\scriptsize ($ \pm $0.28)}}          & \textbf{94.53 {\scriptsize ($ \pm $0.75)}} & \textbf{95.82 {\scriptsize ($ \pm $0.66)}} \\ \hline
	\end{tabular}
\end{table}
\subsubsection {\textbf{CWLK Vs. state-of-the-art malware detectors}}
\label{subsec:comp}
We now compare CWLK based detection with the state-of-the-art Android malware detection solutions to study whether contextual PRG neighborhoods makes good features for malware detection, through experiment E2. For CWLK based detection, ICFG representation with $ h=2 $ is used as it offers the best performance. The Precision, Recall and F-measures of each of these methods are reported in Table \ref{tab:maldetect}. The following observations are made from the table:

\begin{itemize}[leftmargin=*]
\setlength\itemsep{0em}
\item Clearly, CWLK based malware detection outperforms all the compared solutions in terms of F-measure. In particular, our approach outperforms the best performing technique (i.e., \textsc{Drebin}) by 4.87\% F-measure. In terms of Precision, our approach outperforms \textsc{\textsc{Adagio}} and Allix \textit{et al.}'s methods and is comparable to \textsc{Drebin}. In terms of Recall, ours outperforms other methods. 
\item Out of the methods compared, \textsc{Drebin} does not use both structural and contextual features. \textsc{Adagio} and Allix \textit{et al.}'s approaches use structural information but not contextual information. This reveals that capturing both these types of information is the reason for our approach's superior performance, reinforcing our findings from experiment E1.
\end{itemize}

\begin{figure}[t]
	\includegraphics[height=3.6cm,width=9cm]{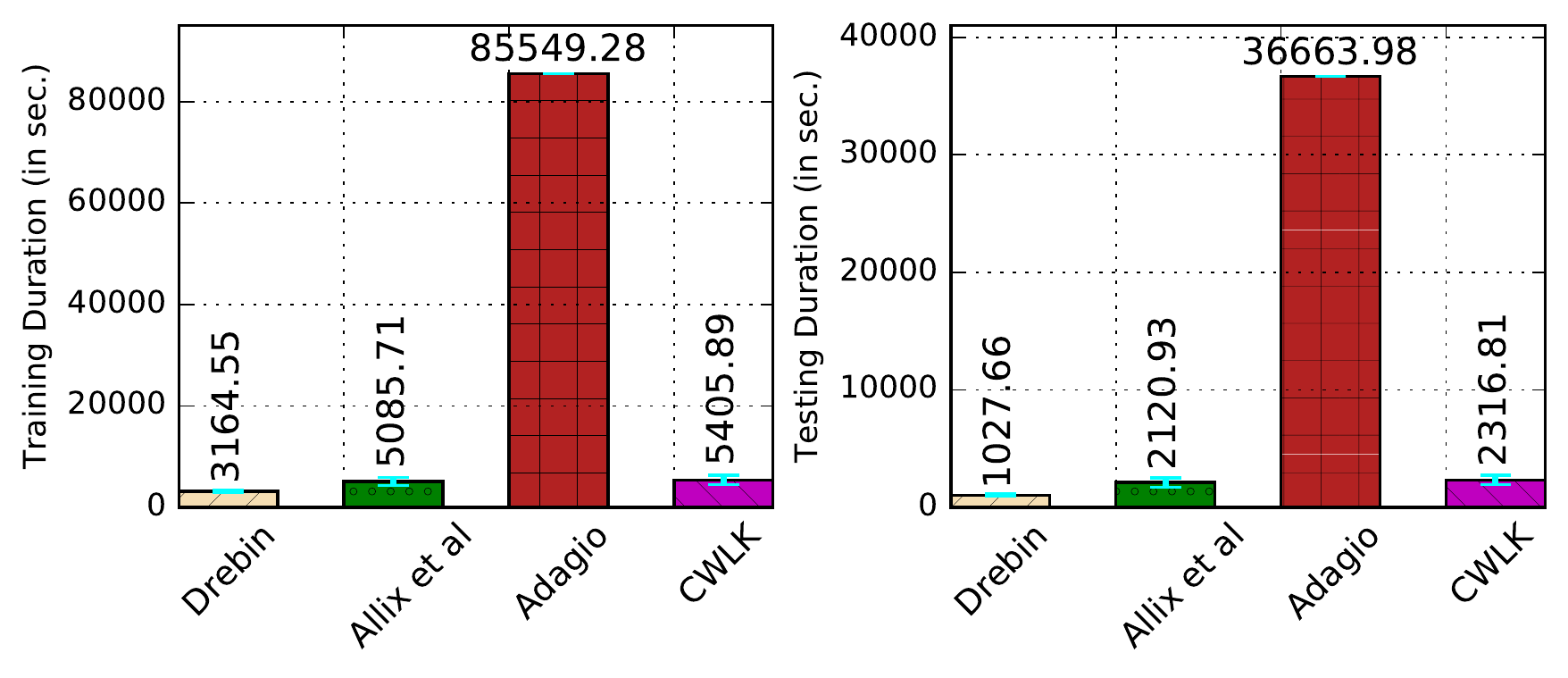}
	\caption{Comparison on Efficiency of CWLK with state-of-the-art malware detectors (avg. over 5 runs) \label {fig:soaeff}}
\end{figure}
\noindent
We now compare the efficiency of CWLK based detection against that of state-of-the-art malware detectors. It is noted that these techniques use different features and classifiers and hence a wide variation in training and testing durations is expected.
The results of this comparison is presented in Fig. \ref{fig:soaeff}, from which the following observations are made:
\begin{itemize}[leftmargin=*]
	\setlength\itemsep{0em}
	\item \textsc{Drebin} being a light-weight non PRG based approach it has significantly higher efficiency than all other methods, including ours.
	\item Allix \textit{et al.}'s method is similar to ours in terms of using PRG based features. Hence our efficiency is comparable to this method.
	\item \textsc{Adagio} uses NHGK and HI kernel SVM in the primal formulation. Hence it takes a prohibitively long time for training and testing. Our method is far more efficient than \textsc{Adagio}.
\end{itemize}
In conclusion, our method's efficiency is comparable to that of other PRG based methods, far better than heavy-weight approaches and inferior to non PRG based light-weight methods.

\noindent
\textbf{Summary.} From experiment E2, we conclude that when compared to state-of-the-art malware detectors, CWLK produces considerably higher accuracy with a practically tractable efficiency, making it suitable for  large-scale real-world malware detection.

\vspace{-2mm}
\section{Conclusion \& Future Work}
\label{sec:conc}
In this paper, we present CWLK, a novel graph kernel that facilitates detecting malware using PRGs. Unlike the existing kernels which capture only the security-sensitive neighborhoods in PRGs, CWLK captures these neighborhoods along with the context under which they are reachable. This makes CWLK more expressive and in turn more accurate than existing kernels. Besides expressiveness, CWLK has two specific advantages: (1) shows high efficiency, (2) supports building explicit feature vector representations of PRGs. 
CWLK is evaluated on a large-scale experiment with more than 50,000 Android apps, and is found to outperform two state-of-the-art graph kernels and three malware detection techniques in terms of F-measure, while maintaining comparable efficiency.\\
\noindent
\textbf{Future work.} In our future work, we plan to investigate incorporating contextual information in other sub-structure based graph kernels such as \cite{NHGK} and \cite{NSPDK} and subsequently, study their suitability for performing malware detection.\\
\noindent
\textbf{Implementation \& Dataset.} We provide an efficient implementation of CWLK and information on the datasets used within this work at: {https://sites.google.com/site/cwlkernel}

\section{Acknowledgment}
\noindent
We thank the authors of \cite{Drebin} and \cite{CSBD}, for their suggestions that helped us re-implement their methods.

\end{document}